**Skip-Level Meetings: Benefits, Risks, and Design Considerations Informed by Bypassed Managers' Experiences**

Samuel Blacher

# Skip-Level Meetings: Benefits, Risks, and Design Considerations Informed by Bypassed Managers' Experiences

Samuel Blacher

This article is based in part on the author's Ph.D. dissertation at Anderson University, Anderson, SC, USA, 2025. (Corresponding author: Samuel Blacher.)

This work involved human subjects in its research. Approval of all ethical and experimental procedures and protocols was granted by the Anderson University Institutional Review Board under Application No. AU062IRB2425.

Samuel Blacher is with Management at Work, Fort Worth, TX 76102 USA (e-mail: Samuel@ManagementAtWork.com).

## Abstract

Skip-level meetings (SLMs), bidirectional, relationship-oriented meetings between employees and managers two or more levels above, are common but not universal, and often recommended in professional outlets. They entail benefits and risks for employees, direct and skip-level managers, and organizations, with multiple factors related to shaping the value-risk balance and appropriate approach. However, SLMs have received limited scholarly attention, particularly regarding bypassed managers' views of that managerial practice. This exploratory qualitative study examined the experiences and perspectives of 25 first-tier software engineering managers from diverse personal and organizational contexts in the multicultural, dynamic, innovation-driven, and high-turnover U.S. tech sector as bypassed managers in SLMs between their direct reports and these reports' second-tier managers. Their accounts, integrated with directly related literature, as well as on intersecting topics such as employee voice, psychological safety, and cultural and generational differences, revealed broad support and preference for exclusion from SLMs, contrary to concerns in prior literature, broad implementation, and high sensitivity to design and contextual factors, such as the significance of follow-up involvement and of individual backgrounds. Findings position SLMs not as a universally beneficial managerial practice, but as context-dependent, requiring deliberate design and implementation, or avoidance. The study extends limited research on SLMs and hierarchy-bypassing interactions, recommends directions for further research, and offers evidence-informed guidance for deciding whether and how to use SLMs to maximize value while mitigating risks.

## Managerial Relevance Statement

Skip-level meetings (SLMs) are a common but non-standard managerial practice with potentially substantial benefits and risks for employees, direct and skip-level managers, and organizations, with multiple factors interacting with them and no "one-size-fits-all" approach. However, they remain sparsely studied, particularly from the perspective of the managers they bypass. This study addresses that consequential research-practice gap through the experiences of 25 first-tier software engineering managers in U.S. tech firms across individual and organizational contexts.

Contrary to concerns suggested by prior literature, participants overwhelmingly supported SLMs and most preferred not to attend their reports' SLMs. The findings further indicate that implementation approach and context matter: for example, exclusion without follow-up involvement was associated with misalignment concerns and negative sentiment, whereas exclusion with follow-up had almost none of both. Psychological safety, leader behavior, preservation of bypassed managers' authority and alignment, meeting format, and cultural, generational, and organizational contexts also emerged as relevant. The study therefore positions SLMs not as a universal “best practice,” but as a context-dependent managerial intervention requiring deliberate decisions about whether to implement them, how to design them, and when to avoid them. It provides evidence-informed guidance for maximizing potential value while mitigating risks and avoiding or addressing pitfalls.

## Introduction

*Cambridge Dictionary* defines “*Skip-level*” (in business) as working “at two levels higher than another person, or a meeting or relationship that involves an employee and a manager at two levels higher” [1]. Some studies [e.g., 2] use it to refer to one or more tiers higher. “Town Halls” or “All Hands” meetings [3], generally used to unilaterally broadcast information on leadership-set topics and rarely evolving into more bidirectional conversations or significantly building or relying on interpersonal skip-level relationships, technically qualify. However, in industry and scholarship, “skip-level meetings” (SLMs) usually refer to smaller, bidirectionally conversational, and relationship-oriented meetings intended to bridge over lower-tier managers [e.g., 4-9]. I focus on that narrower, industry- and scholarly- aligned definition of SLMs.

Managers balance lower- and upper-level agendas, interpret what the top wants, consult downward, and attempt to integrate lower-level concerns upward [10]. SLMs may disrupt that process by directly bridging between the two ends. That link may offer both substantial value to employees, their direct and skip-level managers, and organizations, but also entail significant risks [e.g., 4, 11-14], and interact with personal [e.g., 15-16], organizational [e.g., 17-18], and other contexts relevant to SLMs’ value, risks, and appropriate approach.

SLMs are common, recommended in popular professional outlets, yet far from standard managerial practice [e.g., 4-9, 19]. Despite growing scholarly attention, research on SLMs, employees’ bypassing behavior, skip-level leadership, and the effect of skip-level voice on bypassed managers is scarce [4, 10, 20-21]. Together, inconsistent application, industry interest, limited scholarship, potential rewards and risks, and interacting contexts risk missed opportunities and adverse consequences stemming from intuitive approaches, or their absence.

This exploratory qualitative study, which evolved from PhD dissertation research [22] described in a preliminary, brief "work-in-progress" conference paper [23] (and substantially different from them in content, focus, and structure), addresses the gap in the limited empirical understanding of SLMs, particularly from bypassed managers' perspective, while focusing on the multicultural, dynamic, high-wage, -turnover, and -innovation, and low-unemployment U.S. tech sector [24-32]. By examining the experiences and views of a diverse set of 25 managers across firms, and integrating their accounts with directly related and intersecting literature, the study is guided by the following broad research question while remaining open to emergent themes:

How do first-tier software engineering managers experience and perceive skip-level meetings between their direct reports and these reports' second-tier managers? Particular attention is given to perceived benefits, disadvantages, implementation considerations, and contextual factors that emerge as relevant.

Findings, some of which were unexpected and contrasted with implicit expectations based on prior literature, inform theory and emphasize the need for further research. They also provide a basis for evidence-informed, context-aware practical implementation guidelines.

## Background

### SLMs' Value

Skip-level leaders are considered important sources of authority, idea and initiative support, resources, and upward influence for employees, but information may not always travel up effectively through direct managers, straining communication and risking employees feeling ignored or unable to drive change through traditional channels [2, 33-34]. Bypassing may yield process, morale, and other organizational improvements; favorable employee outcomes, such as raises or transfers; and even improve relationships with direct managers [10, 35]. Skip-level voice aligned with bypassed managers' interests, judgment, and framing may reinforce their upward influence, helping bypassed managers substantiate messages to higher-tier leaders [10].

SLMs may support employees' embeddedness, motivation, and sense of connection to the organization [4] and, building on [20, 36-42], may facilitate higher-tier leaders in enhancing employee access to organizational insights, context, considerations, and constraints that extend beyond direct managers' scope, connect cross-team employees, bring in resources or expertise, and remove barriers, potentially supporting understanding and fostering collaboration, empowerment, trust, open communication, adaptability, software quality, retention, innovation and creativity, performance, and satisfaction. In turn, given that "toxic culture" is a key turnover driver [26], improved understanding of decisions and constraints may support retention.

Skip-level leaders' engagement in meetings with employees in Japanese tech was found to foster employee psychological safety and innovative behavior, as well as lower-level managers' support and sense of accountability [42]. In a tech firm's U.S. sites, employee voice was influenced by beliefs and stories about, interactions with, and behaviors of skip-level leaders [2], suggesting that positive SLM interactions may strengthen voice. In turn, voice may support perceived fairness of decision-making as well as performance [2, 43].

Dynamic capabilities and adaptive leadership research highlight the importance of internal knowledge, lower-tier perspectives, and broad organizational diagnosis while examining political relationships, particularly in dynamic technology environments, to identify opportunities, frictions, gaps in routines and systems, and contextual dimensions [44–46]. SLMs may facilitate such information flows by giving higher-tier leaders more direct, unmediated access to employees' perspectives, experiences, and contexts. In turn, leaders' awareness and clarity about challenges and opportunities, better decision-making, and employees' sense of being heard and valued were associated with motivation, engagement, and satisfaction [13, 47].

Skip-level leaders' attributes and values may trickle down and be associated with performance, trust, cohesion, motivation, integrity, and organizational identification, while direct managers may affect it through their own values [20]. Accordingly, SLMs may promote that effect through direct exposure while reducing lower-level managers' mediator influence.

First-tier software engineering managers may often have weaker leadership skills, which may be more important than their technical skills [48-51]. I suggest that through SLMs, second-tier managers may better identify, prevent, and address concerns early, before they grow, whether first-tier manager-related or not, by soliciting employees' stories and feedback. This may also apply to identifying opportunities. Identified issues may also be leveraged to coach bypassed managers through encountered scenarios and develop their leadership skills.

Where more hierarchical tiers separate employees from top management, employees' perceptions of effective top-down and bottom-up communication with top management may offset their tendency towards a lower sense of attachment, obligation, and commitment to support change [18]. SLMs may be valuable as facilitators of such communication.

Lockheed SLMs were reported to yield recommendations and savings that exceeded the meetings' cost [11-12]. Overall, SLMs may offer direct value for employees, skip-level managers, and even bypassed managers, as well as indirect value, such as through reflecting potential improved team performance, retention, motivation, etc., on direct (bypassed) managers.

**SLMs' Risks**

Managers translate, integrate, contextualize, and coordinate work between their managers and reports [10]. Therefore, through bypassing and direct cross-level links, SLMs may interfere with that function. Prior studies indicated that SLMs may lead to resentment among bypassed managers, tensions, misalignments, mistrust, reduced morale, and hindered authority [11, 13].

A lack of transparency (in this context, SLM-related) may hinder bypassed managers' or employees' psychological safety, while reduced psychological safety may hinder voice and lead to defensive or self-protective behaviors [52]. Managers may perceive bypassing, challenging, or skip-level voice as disloyal, threatening, or norm-breaking, give reports lower performance evaluations, and prefer to hire, maintain, or develop weaker talent [14, 35, 53-54]. Particularly, skip-level voice was associated with sensing a power-status threat and psychological destabilization when bypassed managers do not feel strong, secure, or that they have legitimate authority [14]. Skip-level voice that overrides or contradicts managers' decisions on whether and how an issue should be communicated upward may be destabilizing and hinder the speaker-lower-level manager relationship [10]. Potential managers' job insecurity may lead to engineers' frustration and lower self-efficacy, performance, and organizational citizenship behavior [55].

A "too close" skip-level relationship may lead employees to raise issues with skip-level, rather than direct, managers, especially when their relationship quality with their direct manager is poorer [56], risking misalignment and further relationship deterioration. Additionally, building on [2, 13, 20, 40], mirroring the previously described positive potential amplification of the trickle-down effect of skip-level leaders' values and norms, skip-level interactions may amplify the trickle-down effect of negatively-influencing values and norms, and poor skip-level leader behavior in or related to SLMs may be associated with exhaustion, deviance, reduced creativity, lower employee voice, and more unethical behavior, and, through reduced perceived openness or psychological safety, into demotivation, lower performance, and higher turnover.

Employees' sense of ignored feedback (e.g., in SLMs) may lead to disengagement, lower morale and productivity, and higher turnover, especially in hierarchical settings with limited leadership access, potentially resulting in an employee-leadership disconnect [57]. SLMs are also costly, including stakeholders' time spent preparing, attending, and following up. As described below, lack of consideration and tailoring to various contexts may increase risks and lower value.

**The Significance of Contexts**

Employees' willingness to speak up depends on perceived consequences, impact, their place to speak up, organizational norms, and psychological safety, which matters particularly for knowledge sharing involving interpersonal risk [2, 52, 58]. As such, deliberate management-driven SLMs may shape perceptions of "place to speak up", organizational norms, and consequences differently than ad hoc, uninvited, out-of-norm employee-initiated bypassing. Psychological safety is associated with leaders' transparency and listening, while power distance (though mitigatable through leader behavior) and self-serving behaviors hinder it [52]. Skip-level leaders may impact voice perceptions through behaviors, stories, and interactions, and employees may prefer informal, smaller settings for honest input [2]. Higher psychological safety (relevant for employees and bypassed managers) is associated with reduced defensive or self-protective behavior and may support voice amid power-status differences in SLMs [52].

Perceived job security may support individuals' risk tolerance at work [16], thus may support managers' openness to being bypassed and employees' willingness to speak up proactively and candidly, especially when bypassing direct supervisors. [14] indeed found that direct managers' strength, security, and legitimate authority perception was linked to lower perceived power-status threat or psychological destabilization from reports' skip-level voice. I suggest that low unemployment and high wages among the target population [29-32] may support a sense of security and shape SLM-related perceptions and behaviors. Conversely, where psychological safety or voice are lower, SLMs may add discomfort, concern, or perceived risk, further inhibit voice, or lead to defensive or self-protective behaviors.

Cultural backgrounds may also play a role: employees may hesitate to speak up to leaders, even constructively, for fear that their intentions will be misunderstood because of cultural differences [43]. Cultural norms and values such as hierarchy, power distance, self-protectiveness, uncertainty avoidance, authoritarianism, and micromanagement [59-66], relevant to SLMs, differ across cultures and should be considered. For example, employees from more hierarchical, high power distance, self-protective cultures may be less likely to speak up candidly or proactively to skip-level leaders, especially when bypassing direct supervisors, and managers from more hierarchical, high self-protectiveness and uncertainty avoidance cultures may perceive being bypassed as more norm-breaking, authority-undermining, and distrustful.

This is relevant within the U.S. tech sector: nearly 40% of software developers are foreign-born, mostly in India and China [24], countries characterized by national-level cultural frameworks as significantly different from the U.S. on several workplace-related dimensions, including power distance and hierarchy, uncertainty avoidance, and self-protectiveness (higher scores on these are also associated with authoritarianism and micromanagement) [59-66].

Similarly, preferences relevant to SLMs differ across generations: in general, earlier cohorts may prioritize authority and role clarity, while later cohorts may emphasize participative decision-making and open communication [15]. Accordingly, earlier cohorts may be more likely to see SLMs as norm-breaking and interfering, while more recent cohorts may appreciate them more. The cultural, generational, and other patterns described should be interpreted as contextual lenses rather than deterministic predictors or explainers of individual perceptions or behaviors.

The U.S. tech sector has a high percentage of remote and hybrid jobs [27] and one of the highest turnover rates in the country, allows easy changes, often involving perceptions of hostile, disrespectful, unethical, or noninclusive environments, low satisfaction with supervision, or low identification with the organization [25-26]. These increase risks that require skills to effectively address [17, 67]. As partially mentioned earlier, contrasting leadership skills' importance for first-tier engineering managers, potentially even more than technical skills, they may often be promoted primarily for technical skills and have relatively weaker leadership skills [48-51].

As noted, SLMs may offer particular value in environments where more tiers separate employees from executives. Remote-work environments, or those emphasizing innovation, flexibility, and dynamism, such as the U.S. tech sector, may prefer flatter structures [17, 27-28, 38, 44, 67], where SLMs may better fit organizationally, appeal through potential benefits described earlier (e.g., innovation and adaptability), and be perceived as less norm-breaking.

Conversely, where discipline, synchronization, and strict hierarchy are prioritized, such as in the military [68], or where work is more standardized, well-defined, less dynamic and innovative, or retention is a lower priority, SLMs may offer less value and more risks (e.g., misalignments and friction), as well as deviate more from rooted norms. In summary, numerous personal, organizational, and other contexts interact with SLMs and should be considered.

**Research-Practice Gap and Study Contribution**

SLMs appear common in some environments but absent, even norm-breaking, in others [4-6, 19, 35]. They receive substantial professional attention. For example, an August 2026

Google search I conducted for the exact phrase "skip-level meetings" (with quotation marks) returned "about 71,400 results," including articles in popular outlets [e.g., 7-9].

In contrast, scholarship on SLM and adjacent topics of skip-level leadership, employee hierarchy-bypassing behavior, and the effect of skip-level voice on bypassed managers remains limited [4, 10, 20-21]. To illustrate, an August 2026 Google Scholar search I conducted for the same phrase in article titles returned five results, none from peer-reviewed scholarly journals. Broader searches yielded references to SLMs, though mostly briefly within broader scopes or in conference papers or other non-journal literature. In particular, little is empirically known about bypassed managers' experiences with the managerial practice of deliberate SLMs (rather than issue-specific, ad hoc, employee-initiated manager bypassing). The broad yet non-universal application, industry interest, potential rewards and risks, importance of contexts, and scarce scholarship create risks of misapplication and missed opportunities.

Theoretically, the study extends limited research on skip-level leadership and hierarchy-bypassing interactions by examining SLMs from bypassed managers' perspectives. The findings suggest that established mechanisms such as employee voice and psychological safety may work differently when bypassing is formalized and expected rather than employee-initiated, and that bypassed managers' own psychological safety, authority, alignment, and follow-up involvement may shape how SLMs are experienced. The study further highlights implementation design and contextual factors, including cultural and generational dimensions, as potentially important boundary conditions. These exploratory insights do not establish tested patterns but identify mechanisms and contextual contingencies that may refine theory and guide future research.

Practically, the study promotes awareness of SLMs, their benefits, risks, and interacting factors. Its actionable, evidence-informed insights may help professionals make context-aware decisions about introducing, implementing, or intentionally abstaining from the high-stakes practice to maximize value, minimize risks, and avoid or appropriately address pitfalls.

## Method

### Research Design

This exploratory qualitative study used thematic analysis, informed by interpretive phenomenological principles [69-70], to examine participants' experiences, perceptions, and contexts, deductively and inductively, through a mostly open-ended survey designed to elicit rich responses. Such a survey may benefit the qualitative study [71] and its exploratory nature and allows reaching a broader sample within resource constraints. Additional details about the design and method are described in the Supplemental Material.

### Participants and Sampling

A target sample of 25 first-tier software engineering managers who had served at least a year in their current or recent role at U.S. tech firms with at least 100 employees, varying in personal and organizational contexts, was deemed appropriate (see Supplemental Material) and purposively selected [72] to provide fit, depth, and contextual variation. Smaller firms were excluded because of likely lower hierarchical complexity and closer informal relationships [73-74]. Ages were limited to 25–54 to align with the estimated predominant age range of the target population [75].

Screening considered demographic and organizational diversity but did not ask about SLMs to avoid biasing selection or responses. The sample varied in attributes including firm and team size, age, stakeholder cultural backgrounds, and meeting formats. It generally aligned with some target population estimates, including age groups [75], gender [76], team size [77], and work setting [27]. Other attributes also benefited the study, such as most participants referencing larger firms and permutation availability of SLMs' existence or nonexistence in a current or recent role (see definitions and selected sample attributes in Supplemental Material).

### Data Collection

The online survey used open-ended, conditional logic questions to enhance relevance, engagement, and rich-text input, and avoided leading phrasing. Participants were assured privacy and confidentiality and provided consent (see Supplemental Material). Problematic responses (e.g., partial, flagged as heavily AI-generated) were removed, yielding 25 valid entries.

### Data Analysis

Data were analyzed manually using Braun and Clarke's [69] six-step thematic analysis, with minor adaptations. Interpretive phenomenological principles [70] served as an analytic orientation rather than a separate method and informed attention to holistic consideration of participants' responses within their individual and organizational context, participants' subjective sense-making of their experiences, and reflexive awareness of the researcher's interpretive role. Responses addressing the same issue across different survey questions were considered together within each participant before comparison across participants. Coding and theme development were iterative and both deductive and inductive. This orientation complemented thematic analysis because the research question concerned how managers experienced and perceived SLMs, rather than only identifying recurring cross-case patterns. The sample size supported comparison across multiple contextual permutations while preserving the depth-oriented logic of qualitative thematic analysis.

### Trustworthiness and Rigor

Trustworthiness was supported through complementary sampling, analytic, and external-review procedures. Most participants' LinkedIn profiles were checked to support accurate representation and relevance, and the sample incorporated variation across personal and organizational contexts. During analysis, responses were considered both within participants and across cases; codes and candidate themes were iteratively examined against the underlying responses, contrasting and exceptional cases, related themes, relevant literature, and alternative explanations, and revised accordingly. Variation across participants and environments, together with interpretation through multiple theoretical perspectives, provided complementary forms of triangulation [72, 78].

Before data collection, two members of the dissertation proposal committee reviewed the survey design, which was revised based on their feedback. Subsequently, the doctoral dissertation committee reviewed and provided feedback on the broader research, and a scholarly conference committee reviewed a preliminary work-in-progress paper, providing additional scholarly scrutiny. After the researcher's analysis, an independent doctoral researcher examined the reported findings against the original data and found the interpretations consistent with the material reviewed. These external reviews provided additional perspectives on the study's design and interpretation but were not treated as independent recoding or as establishing inter-coder reliability.

**Researcher Positionality & Reflexivity**

While the author's background as a software engineer and first- and second-tier manager in the tech sector in the U.S. and overseas may have served the study, it also introduced a risk of bias. The author acknowledged this and adhered to reflexivity and rigorous methods during the design, implementation, and interpretation of the research, and also used external reviews.

## Ethical Considerations

The underlying study was approved by Anderson University (South Carolina)'s Institutional Review Board (approval #AU062IRB2425). No identifying information was collected in the survey, and data were described in a non-identifiable manner. Participants reviewed an online informed-consent statement and were required to indicate agreement before proceeding to the survey. No significant ethical or response-bias risk was expected due to the incentive (see Supplemental Material). The author had no pre-existing relationships with participants.

# Results

## Implementation Rates and Overall Sentiment toward SLMs

SLMs were broadly implemented in current roles, and almost all had them in current or recent roles. Strikingly, the sample showed near-unanimous (all but one) conceptual support for SLMs. Furthermore, 83% of those who referenced an implementation valued it, and 13% more favored the concept but disliked the implementation, with some expressing enthusiasm (e.g., "big fan of skip levels;" "grateful that my team has the opportunity to regularly meet with my director"). Those with SLMs in their current but not prior role, or vice versa, favored the concept, and so did both participants who did not have a relevant role with SLMs to reference.

Each participant's overall sentiment was clearly implementation-positive, concept-positive but implementation-negative, or unequivocally negative, with no neutral participants. These results are summarized in Table 1 (percentages were rounded to the nearest whole number):

**Table 1**

*Overall sentiment towards SLMs*

<table>
<tr><th colspan="2" rowspan="3">Subgroup</th><th>Size</th><th colspan="6">Sentiment towards SLMs</th></tr>
<tr><th></th><th colspan="2">Specific positive</th><th colspan="2">Concept-positive</th><th colspan="2">Strong negative</th></tr>
<tr><th>n</th><th>n</th><th>%</th><th>n</th><th>%</th><th>n</th><th>%</th></tr>
<tr><td colspan="2">Current env. with SLMs, no prior env. without</td><td>20</td><td>17</td><td>85%</td><td>2</td><td>10%</td><td>1</td><td>5%</td></tr>
<tr><td colspan="2">Hypothetical (no relevant current or prior SLMs env.)</td><td>2</td><td>N/A</td><td>N/A</td><td>2</td><td>100%</td><td>N/A</td><td>N/A</td></tr>
<tr><td rowspan="3">Allows self-comparison</td><td>Current env. with SLMs, prior without</td><td>1</td><td>1</td><td>100%</td><td>0</td><td>0%</td><td>0</td><td>0%</td></tr>
<tr><td>Current env. without SLMs, prior with</td><td>2</td><td>1</td><td>50%</td><td>1</td><td>50%</td><td>0</td><td>0%</td></tr>
<tr><td>Total allowing self-comparison</td><td>3</td><td>2</td><td>67%</td><td>1</td><td>33%</td><td>0</td><td>0%</td></tr>
<tr><td colspan="2">Total implementation-referencing (non-hypothetical)</td><td>23</td><td>19</td><td>83%</td><td>3</td><td>13%</td><td>1</td><td>4%</td></tr>
<tr><td colspan="2">Grand total: conceptual</td><td>25</td><td colspan="2">Irrelevant</td><td>24</td><td>96%</td><td>1</td><td>4%</td></tr>
</table>

## Perceived Advantages, Disadvantages, and Pitfalls

Perceptions and their contexts largely aligned with literature on relevant theories, such as cultural or generational norms (see below). Generally, positive-sentiment participants described advantages far more than disadvantages, mirrored by negative-sentiment (conditional or not) cases. The sole unconditionally negative participant was also the only one not citing advantages.

Frequently described advantages included open, transparent, two-way, trust-building, and visibility-supporting communication (e.g., “a fantastic outcome is that I have learned things from my boss that [reports] never would've told me. Either feedback on myself as a manager or something about the employee;” “great way to build rapport with the boss’s boss;” reports “get a chance to present themselves, their opinions, and talk about their challenges and accomplishments… building trust and transparency“), alignment and understanding (e.g.,

"understanding of organizational direction, goals, and context of work, constraints, and limitations also improves" from skip-level leader's added context), engagement and motivation (e.g., reports "feel more energized, inspired and motivated"), helping their own leadership and growth (e.g., "I may have blind spots in the way I lead and this process may be able to uncover them;" "help me in building leadership skills by providing context, setting objectives and aiding communication at all levels of the company"), and reports' growth opportunity (e.g., "helps my engineers think bigger picture, how to communicate to an audience one step removed").

Frequently described disadvantages included mixed messaging and misalignments. Repeatedly named pitfalls included earlier-described format aspects, low or altered employee voice, such as due to fear or mistrust, and skip-levels' potentially negatively impacting behavior.

**Participants' Perceptions about Stakeholders' Behaviors**

Sixteen of the 19 participants with positive sentiment toward implemented SLMs expressed little concern about behaviors. They often described second-tier leaders as transparent, empowering, and responsive (e.g., "encouraging participation from the team;" "listens to members' voices;" "incredibly transparent;" "willingly addresses issues… makes it a personal obligation to follow up"), and employees feeling valued and engaged (e.g., "energized from the meetings;" "feel valued that leadership is interested in their opinion;" "appreciate the face time they get with my manager and knowing about certain initiatives directly from him").

Only three of these 19 described negative behaviors, such as superficial feedback handling or employee hesitation. In contrast, all four who referenced SLMs' implementation and had a negative sentiment towards it cited problematic behaviors by their managers (e.g., "rarely respond or follow-up, and change is rarely observed;" regularly reject input superficially such as by "we've always done it this way;" dismisses feedback but shares it as their own) and reports (e.g., “shy to bring up anything… only share things that make them look better;” feel that honestly voicing could hinder their relationship with their supervisor).

**Informal Skip-Level Interactions**

The lack or poor quality of informal skip-level interactions (not part of SLMs) often co-occurred with participants' negative sentiment towards SLMs. In contrast, those who reported informal, valuable interactions tended to have a more positive view of SLMs.

**Format, Alignment with Bypassed Managers, and Misalignment Concerns**

Most reported SLMs were one-on-one and excluded participants, followed by multi-team SLMs, mostly including participants. Interestingly, team-unit SLMs were rarely reported. Participants’ inclusion in SLMs was almost always accompanied by their involvement in preparation and follow-ups. In contrast, those excluded (mostly from one-on-ones) were rarely involved in preparation and inconsistently involved in follow-ups. Only a few expressed concern about preparation, even among those not involved in it.

Surprisingly, most preferred exclusion, often citing reports’ psychological safety, visibility, and opportunity to build relationships with skip-level leaders, as well as self-benefits (e.g., “their time to shine;” “if I join in, members might not talk as much;” “I dramatically prefer it. I get better feedback, my engineers are happier, there's more trust between all three levels”). Additionally, participants included in SLMs substantially more often had negative sentiment towards them than those excluded. However, exclusion without follow-up involvement was often reported with SLM-related misalignment concerns and with negative sentiment.

Remarkably, an exploratory association emerged between inclusion accompanied by preparation and follow-up involvement and informal skip-level interactions (rather than SLM) misalignment concerns. Exclusion accompanied by involvement in follow-ups had the least overall misalignment concerns and negative sentiment reports. The mere existence of informal skip-level interactions was not associated with misalignment concerns or negative sentiment.

Notably, all three who disagreed with their inclusion or exclusion status in SLMs, and preferred the opposite, had implementation-specific negative sentiment. They make three of the four negative-sentiment cases. Two of the three were excluded, and one was included. An exploratory association was identified between such disagreement and stakeholders' cultural backgrounds (see below), hinting at potential mediation. Table 2 summarizes these findings:

**Table 2**

*Participants' inclusion / exclusion, follow-ups, misalignment concerns, and sentiments*

| Overall sentiment towards implementation | n | Included / excluded | Follow-ups involvement | Misalignment concerns due to: SLMs | | Misalignment concerns due to: Informal skip-level interactions | |
|---|---|---|---|---|---|---|---|
| | | | | n | % | n | % |
| Negative | 2 | Excluded | No [a] | 2 | 100% | 0 | 0 |
| | 2 | Included | Yes | 0 | 0 | 1 | 50% |
| Positive | 8 | Excluded | Yes | 1 | 12.5% | 0 | 0 |
| | 5 | Excluded | No | 2 [b] | 40% [b] | 0 | 0 |
| | 6 | Included | Yes | 0 | 0 | 5 | 83.3% |
| N/A (no SLM implementation) | 2 | N/A | | | | | |
| Total | 25 | | | | | | |

[a] One indicated follow-ups existed, but it was unclear whether with their report or manager. [b] Both mentioned a minor concern about misalignment.

When asked what they would change in implementations, format was most frequently cited: participants' inclusion or exclusion, consistency, structure or its absence, preparation and follow-up, the span of employees who have SLMs, and setting (one-on-one, team-unit, etc.).

**Age Cohort Context**

An exploratory association emerged between participants' age groups and their perceptions and sentiments: although the 35–44 age band was the largest, by far, in the sample, it had no negative (conditional or not) sentiment cases, while three out of the four negative (conditional and not) cases were from the smallest age group in the sample, of 45–54, with the fourth coming from the third, last, mid-sized group of participants aged 25-34.

**Cultural Background Context**

While an exploratory association emerged between stakeholders' cultural backgrounds and participants' perceptions, I emphasize that due to the sample's nature, the small sub-groups, and how stakeholders' cultural backgrounds were estimated, this should not be interpreted as a substantiated pattern, but rather as potentially valuable to warrant future investigation. Two groups were identified based on cultural dimensions: I will refer to backgrounds with higher hierarchy, power distance, uncertainty avoidance, and self-protectiveness scores (often associated with authoritarianism and micromanagement) [59-66] as Group A, and those opposite on these dimensional scales as Group B.

Two or three participants (interpretation-dependent, see Supplemental Material: Sample Demographics) out of the four with negative sentiment towards referenced SLMs had Group A managers, and described behaviors generally aligned with such attributes (e.g., directly manage skip-level reports; "quick to dismiss feedback" yet "share it and claim it;" and "not open to criticism or input that could reflect poorly" on them). The fourth and their manager were Group

B, while most of that participant's team was Group A. That participant noted that reports were shy about raising issues except those that made them look good, rarely spoke up unless explicitly asked, and felt that speaking honestly could hinder their relationship with the participant, aligning with Group A dimensions. However, that participant also described their (Group B) manager as unreceptive to feedback and as inhibiting employees' voices.

The only unequivocal negative sentiment case highlighted authority and trust concerns (e.g., "takes away some of my authority and it creates confusion to my direct reports about the hierarchy;" "like my manager doesn't trust me and needs to have some checks;" "presents himself as the main leader"), and was also the only case of both first- and second-tier managers being from Group A. Furthermore, the team was mostly Group B. Conversely, participants largely perceived their Group B managers positively, and in almost all implementation-referring positive sentiment cases, the participant, their manager, and most reports were from Group B.

Participants' disagreement with their exclusion or inclusion in SLMs was also culture-linked: of the three disagreements, the participant included while preferring exclusion was from Group B, and so was their manager, and the two excluded preferring inclusion had a Group A manager. One of these two participants was from Group A, the other was Group B.

## Discussion

The following insights from this early exploration of SLMs should be considered emerging, directional clues rather than tested or deterministic, which may extend theory and guide further research. They may also contribute to practitioners' awareness of the practice, its multi-angled potential value and risks, and interacting factors that may shape the balance between them, with no universal "one-size-fits-all" approach, and support evidence-informed deliberate introduction and implementation of SLMs, or intentional choices to abstain.

## Unexpectedly Broad Implementation and Positive Sentiment

The broad implementation in the sample contrasts with prior sources [4-6, 19, 35]. The broad SLM favoring, particularly among excluded participants, and most participants' preference for exclusion are counterintuitive given implicit expectations based on prior studies [11, 13-14, 53-54]. No relevant source was identified to report implementation rates more specifically in U.S. tech for comparison, or to report bypassed managers' perspectives regarding SLMs. I suggest a few explanations that may translate these apparent contrasts into knowledge extension:

**Sector Norms.** [4, 6, 10, 13-14, 19, 35, 53] did not focus on the sector examined in this study. Remote work, innovation, flexibility, and dynamism may align with flatter structures [17, 27-28, 38, 44, 67], where SLMs may fit better and seem less norm-breaking. [4]'s findings on SLMs' prevalence more broadly in the U.S. suggest that they are substantially less common than in this study's sample. Furthermore, [40]'s finding that, in contrast with India's broader workplace norms, low power distance is prevalent and supports retention in Indian IT organizations, and [42]'s findings on skip-level leaders' engagement value in the tech sector in Japan, another country associated with high power distance [60-61], may further support the notion. Additionally, the sector's "employee market" (high turnover and wages, low unemployment) may lead employers to invest more in employee satisfaction and retention.

**Financial or job security.** High wages and low unemployment rates in the sector [29-32] may have supported participants' risk tolerance, particularly toward skip-level voice [14, 16].

**Cultural norms.** While cultural backgrounds in this study were estimated, and sub-groups were small, along with other sample limitations, I suggest that cultural backgrounds may have played a role, if patterns indeed existed: the sample was U.S.-based, while [5, 13-14, 19] were not; some were conducted in Group A (defined earlier) countries, where bypassing may be less normative. This notion is further supported by culture-related patterns in this study's sample (discussed below).

**Generational norms.** Generational shifts in preferences around authority and role clarity, transparency, and participative decision-making [15] may support more favorable and normative views of SLMs today than in [11, 35, 54]. This notion is further supported by the age-related patterns identified in this study's sample (discussed below).

**Voice context differences.** Some studies [e.g., 10, 13-14, 35, 54, 56] examined employee-initiated, ad hoc, issue-specific skip-level voice. [35] further focused on bypassing to relay organizational dissent, often related to direct managers. I suggest these may be significantly different contexts for their effect on bypassed managers' experiences than skip-level voice delivered through management-initiated SLMs, which may make such voice not only more normative in the organizational culture but also explicitly expected and solicited by management, rather than a specific employee's personal decision. Furthermore, challenging or overriding skip-level voice (likely often in cases of circumvention to express dissent) may be perceived less favorably by bypassed managers [10], while SLMs may cover a broader range of topics.

Overall, findings suggest that SLMs may be highly normative and positively viewed by bypassed managers in some environments or contexts. Positive sentiments dominated across all available permutations of SLMs' existence versus absence in current versus prior roles, as well as among those with no SLM environment to reference (answering conceptually).

### Underlying Perceptions

Positive-sentiment participants generally viewed SLMs as helpful for employees, themselves, and, to some extent, the organization, and described far fewer disadvantages. Negative-sentiment cases largely aligned with negative views of stakeholders' behaviors, as described earlier. This suggests participants may have prioritized "the bigger picture" of reports' interests, the team's outcomes, and their success and growth as leaders over perceived risks.

**Implementation Formats and Misalignment Concerns**

Excluded participants' rare preparation and follow-up involvement suggests missed opportunities to shape SLM agendas, offer context, align on post-meeting insights and actions, or help address and monitor issues. [4]'s findings that, according to surveyed employees, they often did not have a post-SLM follow-up discussion with their direct managers, and based on employees' reports (thus, their knowledge or estimates), SLMs often lacked an agenda altogether, and their direct managers were often not aware of SLMs beforehand, further emphasize the missed opportunities and risks in many implementation approaches.

I suggest explanations for the stark contrast between exclusion (typically in one-on-ones) with and without follow-ups, where exclusion without follow-ups often co-occurred with negative sentiment and SLM-related misalignment concerns, but no informal skip-level interaction misalignment concerns, and when accompanied by follow-ups, misalignment concerns and negative sentiment virtually disappeared: managers' sense of lack of transparency from their leadership may hinder their psychological safety [52]. Exclusion without follow-ups may have been perceived as a lack of transparency by participants' managers. Regardless, a lack of follow-ups, while excluded from SLMs and often also not involved in preparation, may naturally risk misalignment on SLM-discussed issues, disconnect, friction, and hindered authority (perceived or actual). Together, these may translate to negative perceptions.

Paradoxically, inclusion (typically in multi-team SLMs), even with preparation and follow-up involvement, was most accompanied by misalignment concerns, all informal skip-level interactions-related, and negative sentiment. Combined, these may hint at tension between employees' and their managers' psychological safety, sense of security, and voice: because employees may prefer small, informal settings for honest input [2], large forums with direct managers may have suppressed candor and redirected voice to informal interactions, generating respective misalignment concerns, while more private one-on-ones may have offered psychological safety or sense of security that reduced the perceived need to seek alternative voice avenues.

With only one misalignment concern and no negative sentiment cases, exclusion with follow-ups excelled in these aspects. It may have combined employee psychological safety in SLMs and bypassed managers' perceived or actual alignment and psychological safety. Most participants' preference for exclusion, often citing its benefits, may further support this approach.

Nevertheless, inclusion and multi-employee SLMs have advantages, as participants also described, such as enriching SLMs with contexts, multi-angled real-time brainstorming, and direct managers' ability to support employees in speaking up. Other formats, such as three-way (with direct managers), team-unit, or exclusion from multi-team SLMs, were not significantly reported and remain unknown, as does whether inclusion or larger forums were more associated with informal interaction misalignment concerns. Notably, team-unit SLMs' absence suggests missed opportunities to address team context and balance relationships with time cost.

The mere existence of informal (non-SLM) skip-level interactions was not associated with negative sentiment or misalignment concerns on its own, suggesting cautious openness to informal skip-level interactions in relevant environments may be valuable, while ensuring they do not become too frequent and that bypassed managers' authority and alignment are maintained. The exploratory patterns of disagreement with inclusion or exclusion occurring with negative sentiment may indicate the importance of aligning meeting-format expectations.

### Age Cohort Context

Generational preferences may explain the striking perception contrast across age groups: older cohorts may have tended to interpret SLMs as conflicting with role clarity or established authority expectations, whereas younger cohorts may have been more inclined to view them as consistent with participative and transparent management [15]. However, the association was exploratory, and especially given the small subgroups, should not be considered substantiated.

### Cultural Backgrounds

While an exploratory association emerged between stakeholders' cultural backgrounds and participants' perceptions, descriptions of stakeholders' behaviors, and SLM implementation sentiment, I emphasize that due to the sample's nature, the small sub-groups, and how stakeholders' cultural backgrounds were estimated, this finding does not suggest a substantiated pattern, but rather described to justify further investigation. Behavioral descriptions largely aligned with respective cultural dimensions [59-66], such as those with Group A managers (backgrounds associated with higher scores on hierarchy, power distance, self-protectiveness, uncertainty avoidance, authoritarianism, and micromanagement) [59-66] often describing second-tier leaders directly managing reports and low openness to upward feedback, or those with Group B managers (from backgrounds scored lower on these dimensions) often describing their manager as empowering and receptive. These dimensions may also offer context to the co-occurrence of disagreement with inclusion or exclusion, sentiment, and cultural backgrounds, particularly where both excluded participants who preferred inclusion had Group A managers (with one of these two participants also belonging to Group A).

Cultural backgrounds may also play a role through psychological safety and voice lenses: higher power distance is associated with lower psychological safety (mitigatable through leaders' behavior) [52], and employees may hesitate to voice, fearing their intentions may be misunderstood due to cultural differences [43].

**Practical Takeaways**

Given the study's nature, while findings cannot substantiate broad, generalizable implementation guidelines, exploratory practice considerations emerge from them. These may serve as directional clues to inform decisions about whether and how to implement SLMs:

- SLMs are common but non-standard, hold potential value and risks to employees, direct and skip-level managers, and organizations, and interact with individual, organizational, and other contexts shaping value, risk, and whether and how to implement.
- SLMs may be more valuable, normative, supported, and lower risk in more innovative, dynamic, retention-prioritizing, high psychological safety work environments, such as the U.S. tech industry, compared to highly regulated, standardized-work, lower psychological safety work environments, or those prioritizing tight synchronization and strict discipline.
- Cultural and generational contexts should be considered, but as non-deterministic clues.
- Broad implementation and support emerged in the sample. Cultural, generational, and implementation factors, along with disagreement with inclusion or exclusion in SLMs, were associated with misalignment concerns and overall negative sentiment.
- If held, stakeholders' perceptions and behaviors should be monitored, with resentment, friction, misalignments, or other issues promptly addressed (e.g., by coaching, "selling" SLMs, adaptation, and clarifying and ensuring guardrails, such as those below). Ensure:
    - Employees' psychological safety to candidly speak up, with no fear of retaliation, while meaningfully addressing and soliciting feedback.
    - Bypassed managers' psychological safety (such as through ensuring the below).

- Bypassed managers' involvement in preparation and prompt follow-ups to update, discuss, and align on emerging issues after SLMs and informal interactions.
    - Bypassed managers' authority (e.g., skip-level leaders should generally avoid making or communicating decisions that clearly step into bypassed managers' scope or before consulting with them, without significant justification).
    - Bypassed managers' support of the approach, particularly inclusion or exclusion.
    - SLMs do not turn to a default path but remain complementary or for escalation.
    - Accessibility, while not over-accessibility, to informal skip-level interactions.
    - Follow-through on emerging issues and communicating feedback's impact.
- Leaders' behavior in and out of SLMs (e.g., openness, respect, recognition) is crucial and may affect skip-level reports (or bypassed managers) positively or negatively.
- Different formats (e.g., one-on-one, multi-team, inclusion versus exclusion) have their advantages and disadvantages. Small forums that exclude bypassed managers, while maintaining their authority and alignment, may support candid voice and bypassed managers' acceptance. While not examined here and deserving investigation, I suggest considering, if feasible, rotating formats, which may offer added value by combining benefits of different formats.
- SLMs are regularly management-initiated, rather than left for employees to ask for.
- Continuously learn what works well versus what needs improvement. Adapt accordingly.

**Limitations and Future Research**

The findings from this exploratory study of 25 purposively selected participants are not statistically generalizable but offer analytically generalizable insights and directional hints that may inform theory, practice, and future research. Self-selection bias may have existed. Cultural coding was approximate, and some subgroups were small, increasing the risk for some findings.

Larger, representative samples are needed to validate patterns and assess prevalence and transferability. More broadly, the limited scholarly examination of SLMs and the findings, such as contrasting prior studies and highlighting the significance of contexts, warrant further research, including other stakeholders' perspectives, formats, organizational tiers, industries, or countries. SLM-related quantitative investigations, and potential associations with aspects such as performance or retention, are also valuable.

## Acknowledgment

First and foremost, the author sincerely thanks his parents, Yuli and Adela Blecher, for their unwavering support, guidance, and love, which laid the foundation for his academic and professional path. The author also thanks Professor Roxanne Laczo (McCombs School of Business, The University of Texas at Austin) for her extremely valuable and empathetic feedback and support while serving as an external member of his doctoral dissertation committee, as well as Prof. Larry Daniel, Prof. Christopher Rappazinni, and Prof. Clarence Williamson (Anderson University, Anderson, South Carolina) for their helpful guidance, engagement, and support.

## References

[1] Cambridge Dictionary, "Skip-level." Accessed: Aug. 12, 2026. [Online]. Available: https://dictionary.cambridge.org/us/dictionary/english/skip-level

[2] J. R. Detert and L. K. Treviño, "Speaking up to higher-ups: How supervisors and skip-level leaders influence employee voice," *Organ. Sci.*, vol. 21, no. 1, pp. 249–270, Jan./Feb. 2010, doi: 10.1287/orsc.1080.0405.

[3] S. Kjærbeck and M. W. Lundholt, "Communicating strategy in a town hall setting: Is dialog possible?," *Commun. Lang. Work*, vol. 8, no. 1, pp. 19–32, 2022, doi: 10.7146/claw.v8i1.132484.

[4] M. Zebroski, "An examination of skip-level meetings and their relationship with job embeddedness," M.A. thesis, Univ. North Carolina at Charlotte, 2023.

[5] R. Kappal and D. K. Mishra, "Impact of executive isolation on leadership and learning culture: A study across chief executive officers and employees of technology enterprises in India," *Learn. Organ.*, vol. 31, no. 6, pp. 863–883, 2024, doi: 10.1108/TLO-05-2023-0081.

[6] M. S. Neill and S. A. Bowen, "Ethical listening to employees during a pandemic: New approaches, barriers and lessons," *J. Commun. Manag.*, vol. 25, no. 3, pp. 276–297, Jun. 2021, doi: 10.1108/JCOM-09-2020-0103.

[7] S. Nawaz, "How to build a relationship between your employee and your boss," *Harv. Bus. Rev.*, Jan. 27, 2025. Accessed: Aug. 20, 2026. [Online]. Available: https://hbr.org/2025/01/how-to-build-a-relationship-between-your-employee-and-your-boss

[8] P. Falcone, "Preparing for a skip-level meeting," SHRM, Sep. 23, 2021. Accessed: Jun. 9, 2026. [Online]. Available: https://www.shrm.org/topics-tools/news/organizational-employee-development/preparing-skip-level-meeting

[9] S. Tassone, "Skip-level meetings: Finding grains of truth for influence, culture, decision strategy," *Forbes*, Jun. 4, 2026. Accessed: Jun. 9, 2026. [Online]. Available: https://www.forbes.com/councils/forbescoachescouncil/2026/06/04/skip-level-meetings-finding-grains-of-truth-for-influence-culture-decision-strategy/

[10] Z. Jaser, "Voice work, upward influence during change 'when time is of the essence'," *J. Manag. Stud.*, vol. 62, no. 1, pp. 350–380, Jan. 2025, doi: 10.1111/joms.13054.

[11] S. O. Parsons, "The ten commandments for organizational effectiveness," *Proc. Hum. Factors Soc. Annu. Meeting*, vol. 18, no. 3, pp. 325–332, 1974, doi: 10.1177/154193127401800317.

[12] B. J. Kelly, "Results of innovative communication processes on productivity gains in a high technology environment," in *Proc. R&D Productivity: New Challenges for the US Space Program*, 1985, pp. 506-515.

[13] E. Xu, W. Liu, and R. Jia, "The cost to go the extra mile: Consequences of skip-level voice," *Acad. Manag. Proc.*, vol. 2012, no. 1, Art. no. 13525, Jul. 2012, doi: 10.5465/AMBPP.2012.13525abstract.

[14] Y. Yang, J. Li, and T. Sekiguchi, "How supervisors respond to employee voice: An experimental study in China and Japan," *Asian Bus. Manag.*, vol. 20, no. 1, pp. 1–31, Feb. 2021, doi: 10.1057/s41291-019-00075-1

[15] O. Fuchs, E. Lorenz, and L. Fuchs, "Generational differences in attitudes towards work and career: A systematic literature review on the preferences of generations X, Y and Z," *Int. J. Innov. Res. Adv. Stud.*, vol. 11, no. 7, pp. 54–71, Jul. 2024.

[16] M. Sverke, L. Låstad, J. Hellgren, A. Richter, and K. Näswall, "A meta-analysis of job insecurity and employee performance: Testing temporal aspects, rating source, welfare regime, and union density as moderators," *Int. J. Environ. Res. Public Health*, vol. 16, no. 14, Art. no. 2536, 2019, doi: 10.3390/ijerph16142536.

[17] K. Barnes, K. Vione, and Y. Kotera, "Effective leadership practice among senior leaders working from home and in the hybrid workplace across COVID-19," *SN Bus. Econ.*, vol. 4, no. 5, May 2024, Art. no. 55, doi: 10.1007/s43546-024-00651-4.

[18] N. S. Hill, M.-G. Seo, J. H. Kang, and M. S. Taylor, "Building employee commitment to change across organizational levels: The influence of hierarchical distance and direct managers' transformational leadership," *Organ. Sci.*, vol. 23, no. 3, pp. 758–777, May/Jun. 2012, doi: 10.1287/orsc.1110.0662.

[19] COPC Inc., "COPC Inc.'s Global Benchmarking Series 2022," Nov. 2022. Accessed: Aug. 23, 2026. [Online]. Available: https://cx.copc.com/copc-global-benchmarking-series-2022

[20] P. Schilpzand, R. Nerona, Y. Li, S. L. D. Restubog, and Y. Rofcanin, "Leadership in layers: An integrative review on skip-level leadership and an agenda for future research," *Leadersh. Q.*, vol. 37, no. 1, Feb. 2026, Art. no. 101930, doi: 10.1016/j.leaqua.2025.101930.

[21] S. L. Segrest et al., "Leapfrogging at work: Influencing higher levels in the chain of command," *Pers. Rev.*, vol. 49, no. 8, pp. 1769–1786, 2020, doi: 10.1108/PR-07-2019-0382.

**[22]** S. Blacher, "Skip-level meetings: Perspectives of skipped first-tier US tech leaders," Ph.D. dissertation, Anderson Univ., Anderson, SC, USA, 2025.

**[23]** S. Blacher, "Being bypassed: Uncovering skip-level meetings from U.S. frontline technology leaders' perspectives," in *Proc. 21st Eur. Conf. Manage., Leadership, Governance (ECMLG)*, 2025, pp. 494–498, doi: 10.34190/ecmlg.21.1.4176.

[24] American Immigration Council, “Foreign-born STEM workers in the United States,” *American Immigration Council*, Jun. 14, 2022. Accessed: Aug. 19, 2026. [Online]. Available: https://www.americanimmigrationcouncil.org/fact-sheet/foreign-born-stem-workers-united-states/

[25] Y. Li, W. Liu, Q. Qi, and H. Zhu, "Factors and potential solutions of high turnover rates of tech companies and based on mathematical statistics analysis," in *Proc. 2021 Int. Conf. E-Commerce and E-Manag. (ICECEM)*, 2021, pp. 174–179, doi: 10.1109/ICECEM54757.2021.00042.

[26] D. Sull, C. Sull, W. Cipolli, and C. Brighenti, “Why every leader needs to worry about toxic culture,” *MIT Sloan Manag. Rev.*, Mar. 16, 2022. Accessed: Aug. 20, 2026. [Online]. Available: https://sloanreview.mit.edu/article/why-every-leader-needs-to-worry-about-toxic-culture/

[27] Flex Index, "The Flex Report Q4 2024," *Flex Index*, 2024. [Online]. Available: https://www.flexindex.com/report-q4-2024

[28] H. R. Zeidanloo and M. Špaček, “Assessing disruptive potential in the IT sector: A framework for evaluating company-to-company impact,” *J. Innov. Entrep.*, vol. 14, no. 1, Jul. 2025, Art. no. 82, doi: 10.1186/s13731-025-00541-5.

[29] L. Mearian, “Tech unemployment in the US drops to lowest level in more than two years,” *Computerworld*, Jan. 10, 2025. Accessed: Aug. 20, 2026. [Online]. Available: https://www.computerworld.com/article/3800903/tech-unemployment-in-the-us-drops-to-lowest-level-in-more-than-two-years.html

[30] Social Security Administration, "National average wage index," *Social Security Administration*. Accessed: Aug. 19, 2026. [Online]. Available: https://www.ssa.gov/oact/COLA/AWI.html

[31] U.S. Bureau of Labor Statistics, "Civilian unemployment rate," *Charts Related to the Latest "The Employment Situation" News Release*. Accessed: Aug. 19, 2026. [Online]. Available: https://www.bls.gov/charts/employment-situation/civilian-unemployment-rate.htm

[32] U.S. News & World Report, "Software developer salary in 2025: Job outlook & pay," *U.S. News Careers*. Accessed: Jun. 26, 2026. [Online]. Available: https://money.usnews.com/careers/best-jobs/software-developer/salary

[33] J. D. Raelin and C. G. Cataldo, "Whither middle management? Empowering interface and the failure of organizational change," *J. Change Manag.*, vol. 11, no. 4, pp. 481–507, 2011, doi: 10.1080/14697017.2011.630509.

[34] A. Kolk, M. Vock, and W. van Dolen, "Microfoundations of partnerships: Exploring the role of employees in trickle effects," *J. Bus. Ethics*, vol. 135, pp. 19–34, 2016, doi: 10.1007/s10551-015-2727-9.

[35] J. W. Kassing, "Going around the boss: Exploring the consequences of circumvention," *Manage. Commun. Q.*, vol. 21, no. 1, pp. 55–74, Aug. 2007, doi: 10.1177/0893318907302020.

[36] S. M. Nealey and F. E. Fiedler, "Leadership functions of middle managers," *Psychol. Bull.*, vol. 70, no. 5, pp. 313–329, Nov. 1968, doi: 10.1037/h0026505.

[37] K. T. Dirks and D. L. Ferrin, "The role of trust in organizational settings," *Organ. Sci.*, vol. 12, no. 4, pp. 450–467, Jul./Aug. 2001, doi: 10.1287/orsc.12.4.450.10640.

[38] C. S. Mattos, E. Dutra, E. D. Canedo, and G. Santos, “Understanding how power distance affects agile organizations,” in *Proc. 17th Int. Conf. Quality Inf. Commun. Technol. (QUATIC)*, Pisa, Italy, 2024, pp. 311–327, doi: 10.1007/978-3-031-70245-7_22.

[39] S. Oskarsson, T. Svensson, and P. Öberg, “Power, trust, and institutional constraints: Individual level evidence,” *Ration. Soc.*, vol. 21, no. 2, pp. 171–195, May 2009, doi: 10.1177/1043463109103898.

[40] P. Packirisamy, “Managing power distance to retain talent: Evidence from India,” *Int. J. Hum. Cap. Inf. Technol. Prof.*, vol. 8, no. 3, pp. 49–67, Jul. 2017, doi: 10.4018/IJHCITP.2017070104.

[41] G. Tang, Y. Chen, D. van Knippenberg, and B. Yu, “Antecedents and consequences of empowering leadership: Leader power distance, leader perception of team capability, and team innovation,” *J. Organ. Behav.*, vol. 41, no. 6, pp. 551–566, Jul. 2020, doi: 10.1002/job.2449.

[42] M. Matsunaga, "Big boss’s move matters: Impact of the intervention by skip-level leaders on members’ innovative work behaviors and their direct supervisor’s support," in *Proc. ISPIM Innov. Symp.*, 2018, pp. 1-13.

[43] M. Mori, V. Cavaliere, S. Sassetti, and A. Caputo, “Employee voice: A knowledge map to provide conceptual clarity and future research directions,” *J. Manag. Org.*, vol. 30, no. 6, pp. 1873–1899, Nov. 2024, doi: 10.1017/jmo.2022.89.

[44] M. H. G. Schneider, D. K. Kanbach, S. Kraus, and M. Dabić, “Transform me if you can: Leveraging dynamic capabilities to manage digital transformation,” *IEEE Trans. Eng. Manag.*, vol. 71, pp. 9094–9108, 2024, doi: 10.1109/TEM.2023.3319406.

[45] R. A. Heifetz, M. Linsky, and A. Grashow, *The Practice of Adaptive Leadership: Tools and Tactics for Changing Your Organization and the World*. Boston, MA, USA: Harvard Bus. Press, 2009.

[46] F. Yammarino, "Leadership: Past, present, and future," *J. Leadersh. Organ. Stud.*, vol. 20, no. 2, pp. 149–155, May 2013, doi: 10.1177/1548051812471559.

[47] P. Rogers and T. Tierney, "Leadership without control," *Eur. Bus. J.*, vol. 16, no. 2, pp. 78–82, 2004.

[48] R. Garwood, "Supporting the underperforming manager: Teaching the soft skills that can make all the difference," *Hum. Resour. Manag. Int. Dig.*, vol. 20, no. 1, pp. 39–42, 2012, doi: 10.1108/09670731211195963.

[49] E. Kalliamvakou, C. Bird, T. Zimmermann, A. Begel, R. DeLine, and D. M. German, "What makes a great manager of software engineers?," *IEEE Trans. Softw. Eng.*, vol. 45, no. 1, pp. 87–106, 2019, doi: 10.1109/TSE.2017.2768368.

[50] G. P. Sudhakar, A. Farooq, and S. Patnaik, "Soft factors affecting the performance of software development teams," *Team Perform. Manag.*, vol. 17, no. 3/4, pp. 187–205, Jun. 2011, doi: 10.1108/13527591111143718.

[51] A. Coetzee, K. Johnston, and J.-P. Van Belle, "People management skills of information systems managers," in *Proc. 8th Int. Bus. Inf. Manag. Assoc. Conf. (IBIMA)*, 2007, pp. 622–630.

[52] A. C. Edmondson and D. P. Bransby, "Psychological safety comes of age: Observed themes in an established literature," *Annu. Rev. Organ. Psychol. Organ. Behav.*, vol. 10, no. 1, pp. 55–78, Jan. 2023, doi: 10.1146/annurev-orgpsych-120920-055217.

[53] E. R. Burris, “The risks and rewards of speaking up: Managerial responses to employee voice,” *Acad. Manag. J.*, vol. 55, no. 4, pp. 851–875, Aug. 2012, doi: 10.5465/amj.2010.0562.

[54] G. Friebel and M. Raith, “Abuse of authority and hierarchical communication,” *RAND J. Econ.*, vol. 35, no. 2, pp. 224–244, Summer 2004, doi: 10.2307/1593689.

[55] J. Wan, C. Zheng, and L. Wang, “How and when leader job insecurity influences engineer in-role and extra-role behaviors,” *IEEE Trans. Eng. Manag.*, vol. 71, pp. 1891–1899, 2024, doi: 10.1109/TEM.2022.3160690.

[56] W. Liu, S. Tangirala, and R. Ramanujam, “The relational antecedents of voice targeted at different leaders,” *J. Appl. Psychol.*, vol. 98, no. 5, pp. 841–851, Sep. 2013, doi: 10.1037/a0032913.

[57] E. R. Burris, J. R. Detert, and A. C. Romney, “Speaking up vs. being heard: The disagreement around and outcomes of employee voice,” *Organ. Sci.*, vol. 24, no. 1, pp. 22–38, Jan./Feb. 2013, doi: 10.1287/orsc.1110.0732.

[58] J. R. Detert and A. C. Edmondson, “Implicit voice theories: Taken-for-granted rules of self-censorship at work,” *Acad. Manag. J.*, vol. 54, no. 3, pp. 461–488, Jun. 2011, doi: 10.5465/amj.2011.61967925.

[59] GLOBE Project, “2004, 2007 studies,” *GLOBE Project*. Accessed: Aug. 19, 2026. [Online]. Available: https://www.globeproject.com/study_2004_2007?page_id=data

[60] G. Hofstede, "The cultural relativity of organizational practices and theories," *J. Int. Bus. Stud.*, vol. 14, no. 2, pp. 75–89, 1983, doi: 10.1057/palgrave.jibs.8490867.

[61] G. Hofstede, *Culture’s Consequences: Comparing Values, Behaviors, Institutions and Organizations Across Nations*, 2nd ed. Thousand Oaks, CA, USA: Sage, 2001.

[62] M. Castro, “No room for micromanagement in the 21st century,” *Global J. Manag. Bus. Res.*, vol. 20, no. A14, pp. 31–32, 2020.

[63] M. Demmler, R. Ortiz Ayala, and A. Urbiola Solís, "Comparing corporate cultural profiles using the cultural dimensions of Hofstede," *J. Bus. Manag. Sci.*, vol. 6, no. 2, pp. 28–35, 2018, doi: 10.12691/jbms-6-2-1.

[64] T. J. Kaitharath, “The pros and cons of micromanaging: A comprehensive analysis,” *J. Manag. Res. Anal.*, vol. 10, no. 2, pp. 72–73, Jun. 2023, doi: 10.18231/j.jmra.2023.012.

[65] E. Pizzolitto, I. Verna, and M. Venditti, “Authoritarian leadership styles and performance: A systematic literature review and research agenda,” *Manag. Rev. Q.*, vol. 73, no. 2, pp. 841–871, Jun. 2023, doi: 10.1007/s11301-022-00263-y.

[66] M. G. Saenz, T. McGregor, and M. Nguyen, “A cross-cultural examination of the United States, Argentina, and Mexico using Hofstede’s dimensions and the World Values Survey,” *Hum. Soc. Sci. Rev.*, vol. 7, no. 2, pp. 227–236, 2017.

[67] F. Contreras, E. Baykal, and G. Abid, “E-leadership and teleworking in times of COVID-19 and beyond: What we know and where do we go,” *Front. Psychol.*, vol. 11, Dec. 2020, Art. no. 590271, doi: 10.3389/fpsyg.2020.590271.

[68] L. Wong, P. Bliese, and D. McGurk, “Military leadership: A context specific review,” *Leadersh. Q.*, vol. 14, no. 6, pp. 657–692, Dec. 2003, doi: 10.1016/j.leaqua.2003.08.001.

[69] V. Braun and V. Clarke, "Using thematic analysis in psychology," *Qual. Res. Psychol.*, vol. 3, no. 2, pp. 77–101, 2006, doi: 10.1191/1478088706qp063oa.

[70] D. Tuohy, A. Cooney, M. Dowling, K. Murphy, and J. Sixsmith, “An overview of interpretive phenomenology as a research methodology,” *Nurse Res.*, vol. 20, no. 6, pp. 17–20, Jul. 2013, doi: 10.7748/nr2013.07.20.6.17.e315.

[71] V. Braun, V. Clarke, E. Boulton, L. Davey, and C. McEvoy, “The online survey as a qualitative research tool,” *Int. J. Soc. Res. Methodol.*, vol. 24, no. 6, pp. 641–654, 2021, doi: 10.1080/13645579.2020.1805550.

[72] M. Q. Patton, *Qualitative Research & Evaluation Methods*, 3rd ed. Thousand Oaks, CA, USA: Sage, 2002.

[73] O. Mallett and R. Wapshott, “Informality and employment relationships in small firms: Humour, ambiguity and straight-talking,” *Brit. J. Manag.*, vol. 25, no. 1, pp. 118–132, Jan. 2014, doi: 10.1111/j.1467-8551.2012.00836.x.

[74] R. S. Nason, A. McKelvie, and G. T. Lumpkin, “The role of organizational size in the heterogeneous nature of corporate entrepreneurship,” *Small Bus. Econ.*, vol. 45, no. 2, pp. 279–304, Aug. 2015, doi: 10.1007/s11187-015-9632-6.

[75] U.S. Bureau of Labor Statistics, “Employed people by detailed occupation and age,” *Labor Force Statistics from the Current Population Survey*, 2025. Accessed: Jun. 20, 2025. [Online]. Available: https://www.bls.gov/cps/cpsaat11b.htm

[76] Celential.ai, "The percentage of female software engineers in 2023," *Celential.ai*, Mar. 8, 2023. Accessed: Jun. 20, 2025. [Online]. Available: https://www.celential.ai/blog/percentage-of-female-software-engineers/

[77] ICONIQ, “How does your engineering organization stack up? The 2021 ICONIQ Engineering Efficiency Report,” *ICONIQ*, Dec. 8, 2021. Accessed: Aug. 24, 2026. [Online]. Available: https://medium.com/iconiq-growth/how-does-your-engineering-organization-stack-up-the-2021-iconiq-engineering-efficiency-report-b9c08c34250d.

[78] L. A. Guion, D. C. Diehl, and D. McDonald, "Triangulation: Establishing the validity of qualitative studies," *EDIS*, vol. 2011, no. 8, p. 3, 2011, doi: 10.32473/edis-fy394-2011.

## Biosketch

Samuel Blacher received the M.S. degree in engineering and technology management from Oklahoma State University, Stillwater, OK, USA, in 2022, and the Ph.D. degree in leadership from Anderson University, Anderson, SC, USA, in 2025. He has over two decades of technology industry experience across two continents, holding managerial and engineering positions at Fortune 100 companies and startups. His prior roles include Director of Test Engineering at McDonald's, Vice President of Quality Assurance at Mimic Networks, and Senior Quality Assurance Manager at Apple. He currently consults at Management at Work, his technology leadership and organizational culture consultancy. For more information, see: https://www.ManagementAtWork.com